\documentclass{article}
\usepackage{amssymb}
 \newtheorem{Lemma}{Lemma}
 \newtheorem{Proposition}[Lemma]{Proposition}
 
 \newtheorem{Theorem}[Lemma]{Theorem}
 \newtheorem{Conjecture}[Lemma]{Conjecture}

 \newcommand{\Ex}{{\mathbb{E}}}
 \newcommand{\TL}{{\mathbb{T}}}
 \renewcommand{\Pr}{{\mathbb{P}}}

 \newcommand{\ed}{\ \stackrel{d}{=} \ }

 \newcommand{\II}{\mbox{${\mathcal I}$}}

 \newcommand{\EE}{\mbox{${\mathcal E}$}}
 \newcommand{\VV}{\mbox{${\mathcal V}$}}

 \newcommand{\RR}{\mbox{${\mathcal R}$}}

 \renewcommand{\SS}{\mbox{${\mathcal S}$}}
 
 \newcommand{\eps}{\varepsilon}

 \newcommand{\bX}{{\mathbf X}}

 \newcommand{\bx}{{\mathbf x}}

 \newcommand{\sfrac}[2]{{\textstyle\frac{#1}{#2}}}
 
 \newcommand{\len}{\, \mathrm{len}}
 
 \renewcommand{\time}{\mathrm{time}}
 
 \newcommand{\dist}{\mathrm{dist}}
 
 \newcommand{\hop}{\mathrm{hop}}
 \newcommand{\ave}{\mathrm{ave}}
 \newcommand{\area}{\mathrm{area}}
 \renewcommand{\area}{\mathrm{area}}

 \newcommand{\threehex}{\mathrm{3-hex}}

\begin{document}
 \title{Spatial Transportation Networks with Transfer Costs: Asymptotic Optimality of Hub and Spoke Models}
 \author{David Aldous\thanks{Research supported by
 N.S.F Grant DMS-0203062}\\Department of Statistics\\
 367 Evans Hall \#\  3860\\
 U.C. Berkeley CA 94720\\  aldous@stat.berkeley.edu
 \\ www.stat.berkeley.edu/users/aldous
}

 \maketitle

\begin{abstract}
Consider networks on $n$ vertices at average density $1$ per unit area. 
We seek a network that minimizes total length subject to some constraint on journey times, averaged over source-destination pairs.  
Suppose journey times depend on both route-length and number of hops. 
Then for the constraint corresponding to an average of $3$ hops, the length of the optimal network scales as $n^{13/10}$.  
Alternatively, constraining the average number of hops to be  $2$ forces the network length to grow slightly faster than order $n^{3/2}$.  
Finally, if we require the network length to be $O(n)$ then the mean number of hops grows as order 
$\log \log n$.  
Each result is an upper bound in the worst case (of vertex positions), and a lower bound under randomness or equidistribution assumptions. 
The upper bounds arise in simple hub and spoke models, 
which are therefore optimal in an order of magnitude sense.
\end{abstract}

 \vspace{0.1in}

 %{\em MSC 2000 subject classifications.}

 %{\em Key words and phrases.}

 \vspace{0.4in}

 %{\em Short title:}
 %Lengths and costs in networks.

 \newpage
 \section{Introduction}
 \label{sec-INT}
To design a network linking given points (envisage cities) in the plane,
one might specify a cost functional
and a benefit functional on all possible networks,
and then consider the spectrum of networks which are optimal in the sense
of minimizing cost for a given level of benefit.
This suggests a mathematics program:
\begin{quote}
for mathematically simple cost/benefit functionals, study the properties 
(geometry, cost and benefit values)
of such optimal networks as the number $n$ of points tends to infinity, 
under either worst-case or typical case point positions.
\end{quote}
Network design problems arise in many applied fields, 
but real-world modelling leads to more complicated functionals than we have in mind 
\cite{yang-bell}.
Algorithmic issues in network optimization are another 
well-studied topic \cite{gendron98multicommodity}, 
and the flavor of ongoing research in one academic Engineering community can be seen by browsing the journal {\em Transportation Research}.
But the combination of $n \to \infty$ asymptotics and 
simple functionals has apparently not been systematically studied, 
though some recent statistical physics literature 
touches upon similar ideas
(in particular
\cite{gastner-2006-74},  discussed in section \ref{sec-discuss}).

Write 
$\bx = \bx^n = \{x_1,x_2,\ldots,x_n\}$
for a configuration of $n$ vertices in the square 
$[0,n^{1/2}]^2$ 
of area $n$ 
(we find this ``density $1$" scaling convention more natural than others).
So $x_i$ is the position of vertex $i$.
  Create a connected {\em network}  $G_n$ by adding edges:
 edges are line-segments with their natural Euclidean lengths,
and edges may meet at junctions not in the given vertex-set $\bx$.
Define the network length $\len (G_n)$ as the sum of the edge-lengths.  
This is perhaps the simplest choice of functional to serve as ``cost".  
The kind of ``benefit" we have in  mind is that the network enables material to be transported between typical vertices $i,j$ in short time. 
The simplest formalization of ``time" is as route-length, and the trade-off in this setting is studied in \cite{me116}. 
In this paper we envisage a type of  setting exemplifed by passenger airline travel where direct flights are not always available, and by package shipments via commercial carrier.  
In such settings, the time to transfer between successive stages (``hops") of the journey is of the same order of magnitude as the transit times.  
As the simplest model for this phenomenon, 
fix a parameter $\Delta > 0$.
For any route $\pi$ consisting of successive edges $e_1,e_2,\ldots,e_m$ define the travel
time for this route to be
\[ \time(\pi) = \Delta (m-1) +  n^{-1/2} \sum_i \len(e_i) .\]
Here $\len(e_i)$ denotes the length of edge $e_i$.  The first term represents time to transfer \
between successive hops, and the second term represents transit time.   
Note the scaling is chosen to make both terms have the same order of magnitude, as $n \to \infty$, 
assuming the number of hops stays bounded.
Now define the journey time between any pair of vertices $(i,j)$ to be
\[ \time(i,j) = \min \{\time(\pi): \ \pi \mbox{ a route from $i$ to $j$} \} \]
and then for the network $G_n$ define the average journey time statistic
\[ \time(G_n) = \ave_{(i,j)} \time(i,j) . \] 
Adding extra edges to a network will decrease $\time(G_n)$ at the expense of increasing $\len(G_n)$.
In this paper we study the trade-off between $\len(G_n)$ and $\time(G_n)$ as $n \to \infty$.

Recall the concept of a {\em hub and spoke} network
\cite{wiki:spoke-hub}, 
which became familiar in the U.S. in the 1970s with the FedEx (overnight package delivery) hub at Memphis, and in the 1980s with Delta Airlines establishing hubs in Atlanta, New York, Cincinnati and Salt Lake City.  
In such networks, a typical vertex is linked to only one or two relatively nearby hubs, while the hubs are well-connected to each other.
Intuition (or the fact they form the basis of multibillion dollar industries) suggests that such networks 
should be roughly optimal with respect to the trade-off between length and time as defined above. 
We will prove asymptotic results in this spirit.  
Essentially, our results say that simple hub-and-spoke networks can achieve lengths which are the same order of magnitude as the optimal network.  
Understanding properties of the exact optimal networks seems a much harder problem, which we do not tackle.

\subsection{Some mathematical set-up}
\label{sec-setup}
In studying this topic we found it helpful to first consider the {\em random model} 
\begin{equation}
\mbox{
the $n$ vertices of $\bX^n$ are random (uniform, independent) in the square $[0,n^{1/2}]^2$
}
\end{equation}
and then consider the ``worst case", i.e. seek upper bounds for an arbitrary configuration $\bx^n$,
and finally consider lower bounds, which inevitably require some constraints making the configuration 
$\bx^n$ be fully two-dimensional.
It turns out that the order of magnitude for the random model is determined by 
the general upper and lower bounds,
so we haven't recorded results for the random model explicitly 
except in Theorem \ref{T1} where a natural ``sharp optimality" conjecture arises.

For lower bounds we will use a quantitative version of an equidistribution property.  
Take integers $L^\prime_n$ and consider the natural partition of the square
$[0,n^{1/2}]^2$ into subsquares $\sigma$ of side $s_n = n^{1/2}/L^\prime_n$.
For a configuration $\bx^n$,
define the 
``smoothed empirical distribution" 
$\psi$ as the distribution of a position $Y$ obtained by picking a uniform random 
point $x$ from $\bx^n$ and then taking $Y$ uniform on the subsquare containing $x$. 
Write $D_{s_n}(\bx^n)$ for variation distance between $\psi$ and the uniform distribution on 
$[0,n^{1/2}]^2$.  
Explicitly,
\[ D_{s_n}(\bx^n) = % for variation distance between $\psi$ and the uniform distribution on 
\frac{1}{2n}
\sum_\sigma \big| 
\area(\sigma) - 
|\bx^n \cap \sigma| 
\big| \]
where the sum is over the subsquares $\sigma$ 
and where 
$|\bx^n \cap \sigma| $
is the number of points of $\bx^n$ that lie in the subsquare $\sigma$.
See section \ref{sec-equi} for remarks on this definition.

Define
\[ c_1 = \mbox{ mean distance between two random points in the unit square}
\approx 0.52 . \]
Write $d(\cdot,\cdot)$ for Euclidean distance, and write 
\[  \dist(\bx) = n^{-1/2} \ave_{(i,j)} d(x_i,x_j) \]
for normalized average inter-vertex distance.  
For any network $G_n$ on vertices $\bx^n$ we have 
\[ \time(G_n) \geq \dist(\bx^n) \]
and so in comparing networks on different vertex-sets it is natural to consider the difference 
$\time(G_n)  -  \dist(\bx^n)$.
It is clear that in the random model, the expectation satisfies
\begin{equation}
 \Ex \ \dist(\bX^n) \to c_1 .
\label{c-ave}
\end{equation}

Finally, define the {\em hop-diameter} of a network to be the maximum (over $i$ and $j$) 
of the minimum number of hops in a route from $i$ to $j$.

\subsection{Statement of results}
We start with what turns out to be the most interesting point on the spectrum, where the constraint on ``time" corresponds to an average of three steps in a route, and we seek to minimize network length subject to that constraint.
Let us first consider a specific network.  
Write $\TL$ for the vertex-set of the  {\em triangular lattice} whose Voronoi tessallation partitions the plane into hexagonal cells; standardize so that the density of vertices (and hence the area of hexagons) 
equals $1$.
It is classical \cite{MR0165423} that this is the ``best" tessallation in terms of mean distance 
from a random point to the closest lattice vertex: 
see \cite{MR1903153} for a recent treatment and history, and see \cite{MR1809149} for a broad account of optimality properties of the triangular lattice.
Define numerical constants
\[c_2 = \mbox{mean distance from center of unit-area hexagon to random point in hexagon} 
\approx 0.377\] 
\[c_3 = \mbox {distance from center of unit-area hexagon to corner of hexagon boundary} 
= \sfrac{3}{2} c_2.\]

{\bf The 3-hop hexagonal network.}
Write $\TL (L,n)$ for the vertices of the triangular lattice after scaling by $L$, so that cells have area $L^2$, and then restricting to vertices whose cells intersect the square 
$[0,n^{1/2}]^2$.
Define the network 
$G^{\threehex}(\bx^n,L)$ 
to have edges \\
(i) between each pair of vertices of $\TL(L,n)$;\\
(ii) from each vertex $x_i$ to the closest vertex of $\TL(L,n)$.
\\
So this is a ``hub and spoke" network where the hubs are the vertices
of $\TL(L,n)$; obviously it has hop-diameter $3$.
\begin{Theorem}
\label{T1}
(a) In the random model, choosing
\[
L_n = (2c_1/c_2)^{1/5} n^{3/10}
\]
we have 
\begin{eqnarray}
\Ex\  \time(
G^{\threehex}(\bX^n,L_n)
) &\to& 2 \Delta + c_1. 
\label{3-time}\\
\Ex \ \len(
G^{\threehex}(\bX^n,L_n)
) &\sim &
\beta n^{13/10} ; \quad 
\beta := 5 \cdot 2^{-9/5}c_1^{1/5} c_2^{4/5} .
\label{3-len}
\end{eqnarray}
(b) For arbitrary $(\bx^n)$, 
with the same choice of $L_n$ 
\begin{eqnarray}
 \time(
G^{\threehex}(\bx^n,L_n)) - \dist(\bx^n) 
 &\leq& 2 \Delta + o(1). 
\label{3-time-wc}\\
\len(
G^{\threehex}(\bx^n,L_n)
) &= &
O( n^{13/10} )  .
\label{3-len-wc}
\end{eqnarray}
(c) Suppose $(\bx^n)$ satisfies the equidistribution property
\[ D_{s_n} \to 0 \mbox{ for some } s_n = O(n^{3/10}) . \] 
Then any networks $G_n$ on $\bx^n$  satisfying 
 \begin{equation}
  \time(G_n)
- \dist(\bx^n) 
 \leq 2 \Delta + o(1) \label{2+}
 \end{equation}
 must satisfy 
 \[ \len(G_n) = \Omega(n^{13/10}) , \] 
 and this conclusion remains true if (\ref{2+}) is replaced by the weaker assumption 
  \begin{equation}
  \time(G_n)
- \dist(\bx^n) 
 \leq (3 - \eps) \Delta + o(1) \label{3-}
 \end{equation}
 for fixed $\eps > 0$.
 \end{Theorem}
(Recall that $a_n = \Omega(b_n)$ means $b_n = O(a_n)$).
In words, this result says that for networks with three-hop routes (i.e. an average of three hops), this particular 
hub and spoke model has length of order $n^{13/10}$ in both the random setting and the worst-case setting; and no other network can improve this order of magnitude 
(assuming the equidistribution property) without increasing the average number of hops to almost $4$.  
The unusual scaling exponent $13/10$  makes the result memorable. 
The proof (section \ref{sec-p1}) uses nothing more than freshman calculus for (a) and (b), and little more than 
the coupling interpretation of variation distance for (c). 
We conjecture that the constant in (\ref{3-len}) is optimal: 
\begin{Conjecture}
In the random model, any networks $(G_n)$ satisfying 
\[ \Ex \  \time(G_n)
\leq 2 \Delta + c_1 + o(1) \]
must satisfy
\[ \Ex \len(G_n) \geq (\beta - o(1)) n^{13/10}  \] 
for $\beta$ defined at (\ref{3-len}).
\end{Conjecture}
This may be fussy to prove, because it obviously involves the optimality property
of the triangular lattice.  Incidently, the only reason we use the triangular lattice at all is to formulate this conjecture; for order-of-magnitude results the square lattice would work just as well.

Moving along the trade-off spectrum, the next result says that, to 
decrease the average number of hops to $3 - \eps$,  the length of the network must increase from 
order $n^{13/10}$ to order $n^{3/2}$; and with this length we can decrease the average number of hops to $2 + \eps$.
\begin{Proposition}
\label{P2} 
For $ 1 > \eps>0$ there exist constants 
$0 < b(\Delta,\eps)$ and  $B(\Delta,\eps) < \infty$
satisfying the following.\\
(a) For arbitrary $(\bx^n)$ there exist  networks $(G_n)$ such that
  \begin{eqnarray}
  \time(G_n)
- \dist(\bx^n) 
 &\leq &(1 + \eps) \Delta + o(1) \label{1+}\\
  \len(G_n) &\leq& B(\Delta,\eps) n^{3/2} 
 \end{eqnarray}
 and we may choose such networks to have hop-diameter $2$. \\
(b) Suppose $(\bx^n)$ satisfies the equidistribution property
\[ D_{s_n} \to 0 \mbox{ for some } s_n = o(n^{1/2}) . \] 
Then any networks $G_n$ on $\bx^n$  satisfying 
 \begin{equation}
  \time(G_n)
- \dist(\bx^n) 
 \leq (2 - \eps) \Delta + o(1) \label{2-}
 \end{equation}
 must satisfy 
 \[ \len(G_n) \geq (b(\Delta,\eps) - o(1)) n^{3/2} .\] 
\end{Proposition}
The proof (section \ref{sec-p2}) is just a simpler analog of the proof of Theorem \ref{T1}.
The construction in (a) is again a hub-and-spoke network, but now each point is linked to each hub.

The situation at ``$2$ hops" is somewhat different from the situation at ``$3$ hops", as the next result shows: 
the optimal length grows just faster than order $n^{3/2}$.
See section \ref{sec-p3} for the proof.
\begin{Proposition}
\label{P3} 
(a) For arbitrary $(\bx^n)$ and for $\omega_n \to \infty$ arbitrarily slowly, there exist networks $(G_n)$ such that
  \begin{eqnarray}
  \time(G_n)
- \dist(\bx^n) 
 &\leq & \Delta + o(1) \label{2}\\
  \len(G_n) &\leq& \omega_n n^{3/2} 
 \end{eqnarray}
 and we may choose such networks to have hop-diameter $2$. \\
(b) Suppose $(\bx^n)$ satisfies the equidistribution property
\[ D_{s_n} \to 0 \mbox{ for some } s_n = o(n^{1/2}) . \] 
Then any networks $G_n$ on $\bx^n$  satisfying 
 \begin{equation}
  \time(G_n)
- \dist(\bx^n) 
 \leq  \Delta + o(1) \label{1-}
 \end{equation}
 must satisfy 
\[ n^{-3/2} \len(G_n) \to \infty . \]
\end{Proposition}
It is easy to guess intuitively how Theorem \ref{T1} and Proposition \ref{P3} should extend to the 
``average $m$-hops" setting for general $m \geq 4$.  
We outline this in section \ref{sec-mhop} without details.

 For the final result we look at the trade-off the other way round. 
 Constraining network length to be $O(n)$, how small can we make the average number of hops?
 Precisely, for vertices $i,j$ in a network $G_n$ define 
 $\hop(i,j)$ as the smallest number of hops in a route from $i$ to $j$, 
 then define
 $\hop(G_n) = \ave_{(i,j)} \hop(i,j) \leq \mathrm{hop-diameter}(G_n)$.
 
\begin{Theorem}
 \label{T4}
 (a) For arbitrary $(\bx^n)$ there exist networks $(G_n)$ such that 
 $\len(G_n) = O(n)$ and 
 $\hop(G_n)  % O(\log \log n)$.\\
\leq \mathrm{hop-diameter}(G_n) 
\leq (2 + o(1)) \log_2 \log n$. \\
 (b) Suppose $(\bx^n)$ satisfies the equidistribution property
\[ D_{s_n}(\bx^n) \to 0 \mbox{ for some } s_n = % o(). \] 
o(\exp(\log^\beta n)) \]
for some $\beta > 0$.
Then any networks $G_n$ on $\bx^n$  satisfying $\len(G_n) = O(n)$ 
must satisfy  $\hop(G_n)  % \Omega(\log \log n)$.
\geq (1 - \beta - o(1)) \log_2 \log n$.
\end{Theorem}
\begin{Conjecture}
 The lower bound in (b) can be improved to 
$2(1 - \beta - o(1)) \log_2 \log n$. 
\end{Conjecture} 
The construction for (a) uses a hierarchical network.
See section \ref{sec-PT4} for the proof and a remark relating to the conjecture.

\subsection{Discussion}
\label{sec-discuss}
There is a quite extensive literature on heuristic algorithms for designing optimal
hub and spoke type networks: see the survey \cite{Bryan-May}.
For a typical relevant paper in the discipline of transportation research (the optimal network for an airline in Western Europe)
see \cite{adler-ber}.
The general phenomenon expressed by our results -- that there is a big difference between 
average 3 hop and average 2 hop networks -- has of course been noted in this literature 
(e.g. \cite{yoon-yoon}), being very visible in any moderately large example,
but we haven't found any previous attempt at theoretical analysis of this difference.
As well as the specific scaling exponents we have derived, 
our work serves to verify that natural hub and spoke networks are indeed optimal in an order of magnitude sense.

Closest to our work is that of
Gastner and Newman \cite{gastner-2006-74}, who describe essentially the same model. 
Without giving theoretical analysis, they show
(their Figure 5) pictures of optimal networks for the U.S. 
(discretized to $200$ points and population-weighted).  
Their figure illustrates the expected qualitative behavior as $\Delta$ increases: fewer hubs and longer spokes.  It would be interesting to quantify this behavior. 
Unfortunately, as illustrated schematically in Figure 1, 
 our results are insensitive to the value of $\Delta$.
Of course this highlights a general point that $n \to \infty$ asymptotics may not capture the features
of real-world relevance.  
Note that $\time(G_n)$ involves the product of $\Delta$ and average hop number.
Varying $\Delta$ while holding $\time(G_n)$ fixed, as in \cite{gastner-2006-74}, 
is moving along a hyperbola in Figure 1, so one expects qualitative changes in the geometry of the optimal network as one passes through integer values of average hop number.

Implicit in our results is a remarkable and counter-intuitive fact.  
Consider first the $\Delta = 0$ case (no transfer cost).  
Some length of network is required to make the network connected, and one might guess that 
a non-negligible extra length of total network would be required to ensure that average route-length between two cities was only $(1 + o(1))$ times average Euclidean distance.  
But as studied in \cite{me116}, one can achieve this with total network length of only $(1 + o(1))$ times the length of the shortest connected network.  
Analogously, in the present  $\Delta > 0$ setting, although $\time(G_n)$ depends on both route-length and number of transfers, making the route-length be only negligibly longer than Euclidean distance is easy to achieve, so our constraint on $\time(G_n)$ is effectively just a constraint on the average number of transfers.

\setlength{\unitlength}{0.6in}
\begin{picture}(4,4)(-1.5,0)
\put(0,0){\line(1,0){3.5}}
\put(0,0){\line(0,1){3.5}}
\put(1,0.4){$n^{5/2}$}
\put(1,1.4){$n^{3/2}$}
\put(1,2.4){$n^{13/10}$}
\put(1,3.4){$n^{7/6}$}
\put(0,1){\line(1,0){1.1}}
\put(1.3,0.92){$n^{3/2} \omega_n$}
\put(2.2,1){\line(1,0){1.3}}
\put(-0.3,-0.1){1}
\put(-0.3,0.9){2}
\put(-0.3,1.9){3}
\put(-0.3,2.9){4}
\put(0,2){\line(1,0){1.1}}
\put(1.3,1.92){$n^{13/10} $}
\put(2.0,2){\line(1,0){1.5}}
\put(0,3){\line(1,0){1.1}}
\put(1.3,2.92){$n^{7/6} \omega_n$}
\put(2.2,3){\line(1,0){1.3}}
\put(2.2,-0.3){$\Delta$}
\put(-1.6,1.73){ave. hop}
\put(-1.6,1.4){number}
\end{picture}

\vspace{0.2in}
{\bf Figure 1.}
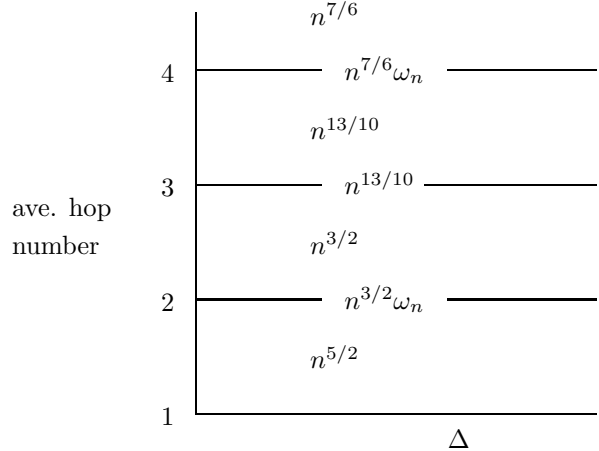
{\small Schematic for order of magnitude of length of shortest network, as a function of 
the ``transfer cost" parameter $\Delta$ (horizontal scale) and the average number of hops (vertical scale).  This illustrates that our results are insensitive to the value of $\Delta$; it would be interesting to find more refined results capturing the influence of $\Delta$.
}

%\newpage
\section{Proofs}
\subsection{On the equidistribution hypotheses}
\label{sec-equi}
Proofs of lower bounds use the following coupling inequality, which is an immediate consequence of 
the definition (section \ref{sec-setup}) of $D_{s_n}(\bx^n)$ and the coupling interpretation (\cite{bhj92} (A.1.4)) 
of variation distance.
\begin{Lemma}[coupling inequality]
\label{L-couple}
Let $\bx = \bx^n$ be an arbitrary configuration.
Let $I$ be uniform on $\{1,2,\ldots,n\}$ and let $U_n$ be a uniform point in 
$[0,n^{1/2}]^2$. 
Then we can couple $x_I$ and $U_n$ such that
\[ \Pr(d(x_I,U_n) > s_n \sqrt{2}) \leq D_{s_n}(\bx) . \] 
\end{Lemma}
Lower bounds use hypotheses of the format
\begin{equation}
D_{s_n} \to 0  \label{Dn2}
 \end{equation} 
for  $(s_n)$ satisfying some bound.    
 The classical equidistribution property
 \begin{quote}
 the empirical distribution of $\{n^{-1/2}x^n_i, 1 \leq i \leq n\}$ converges weakly to the uniform
 distribution on $[0,1]^2$
 \end{quote}
 is equivalent to the property (\ref{Dn2}) holding for {\em some} $s_n = o(n^{1/2})$: 
 this is the hypothesis in Propositions \ref{P2} and \ref{P3}. 
 Replacing one sequence $(s_n)$ by a slower-growing sequence makes (\ref{Dn2}) a 
 stronger assumption, so the hypotheses in Theorems \ref{T1} and \ref{T4} are stronger.  
 A configuration of independent uniform random points satisfies (with probability $\to 1$) 
the requirement (\ref{Dn2}) for any specified sequence $s_n \to \infty$, however slowly.  
So for the random model our results establish the correct order of magnitude in the settings considered.

\subsection{Proof of Theorem \ref{T1}}
\label{sec-p1}
We first prove parts (a) and (b).  
Consider an arbitrary configuration $\bx^n$.
As a route in $G^{\threehex}(\bx^n,L_n)$ between $i$ and $j$, use the  $3$-hop route 
$i \to v(i) \to v(j) \to j$
where $v(i)$ is the vertex of $\TL (L,n)$  closest to vertex $i$. 
For this route $\pi$
\[ \time(\pi) \leq 2\Delta + n^{-1/2}(d(i,j) + 4c_3 L_n) . \] 
So for any choice of $L_n = o(n^{1/2})$ we get 
\[ \time(G^{\threehex}(\bx^n,L_n)) \leq 2\Delta + \dist(\bx^n) + o(1) . \]
This implies the time-bound (\ref{3-time-wc}) in the worst case, and 
(using (\ref{c-ave})) establishes (\ref{3-time}) in the random model.

To calculate network length, note there are $N_n \sim 
n/L_n^2$ 
hexagonal cells intersecting the square of area $n$.
So the total length of edges linking cell centers is asymptotic to 
\[ {N_n \choose 2} \times c_1 n^{1/2} \sim  \sfrac{1}{2} (n/L_n^2)^2 \times c_1 n^{1/2} . \]
In the random model,
the mean distance from a vertex to the center of its hexagon is 
$c_2 L_n$, so (the effect of boundary hexagons being asymptotically negligible) the mean total length of such edges is $\sim c_2 L_n n$.  
Similarly in the worst case, the total length of such edges is bounded by $c_3L_n n$.
Combining: in the random model
\[
\Ex \len(
G^{\threehex}(\bX^n,L_n) 
) \sim 
2^{-1}c_1 L_n^{-4}n^{5/2} + c_2 L_n n . \]
Minimizing the right side over $L_n$, the minimum is attained by 
\[
L_n = (2c_1/c_2)^{1/5} n^{3/10}
\]
and the minimized value is
\[
\beta n^{13/10} ; \quad 
\beta := 5 \cdot 2^{-9/5}c_1^{1/5} c_2^{4/5} . \]
This gives (\ref{3-len}), and the  worst-case bound (\ref{3-len-wc}) is similar.

The proof of (c) is based on the following lemma.
Fix a network $G_n$ on vertices $\bx^n$, and let $D_{s_n} = D_{s_n}(\bx^n)$ be as in 
section \ref{sec-setup}.
Let $I$ and $J$ be independent uniform random vertices from $\{1,2,\ldots,n\}$.
Consider the minimum-time route between $I$ and $J$.
Write $p_n(a,b)$ for the probability that this route has exactly $3$ hops, 
the middle hop having length $\geq b$ 
and the other hops having lengths $\leq a$.
\begin{Lemma}
\label{L1}
\begin{eqnarray}
p_n(a,b) &\leq&
2D_{s_n} + 
\frac{2\pi^2(a+s_n \sqrt{2})^4 
\ \len(G_n)}
{n^2 b}
\label{pnupper} \\
1-p_n(a,b) &\leq&
2D_{s_n} + 
\frac{\pi (2a+b+2^{3/2}s_n)^2}{n}
+ \frac{4 \ \len(G_n)}{an} 
+\Pr(H_n \geq 4) 
\hspace*{0.5in} 
\label{pnlower} 
\end{eqnarray}
where $H_n$ is the number of hops on the route from $I$ to $J$.
\end{Lemma}
\begin{proof}
By  the coupling inequality we can couple $x_I$ and $x_J$ to independent uniform points 
$U_1,U_2$ in  $[0,n^{1/2}]^2$ 
so that
\[ \Pr(d(x_I,U_1) > s_n \sqrt{2} \mbox{ or } d(x_J,U_2) > s_n \sqrt{2} ) \leq 2D_{s_n} . \] 
From the definition of $p_n(a,b)$,
\[ p_n(a,b) \leq 2D_{s_n} +
\Pr(\exists e = (\hat{v}_1,\hat{v}_2) \in \EE_b: \ 
d(U_r,\hat{v}_r) \leq a + s_n\sqrt{2}, r = 1,2)  \] 
where $\EE_b$ denotes the set of edges of $G_n$ with length $\geq b$, regarding each such edge as two directed edges.   
For a particular such directed edge $(\hat{v}_1,\hat{v}_2)$,
\[ \Pr(d(U_r,\hat{v}_r) \leq a + s_n\sqrt{2}, r = 1,2) \leq \left(\frac{\pi (a+s_n\sqrt{2})^2}{n}\right)^2 \] 
and so 
\[ p_n(a,b) \leq 2D_{s_n} + 
|\EE_b | \times \left(\frac{\pi (a+s_n\sqrt{2})^2}{n}\right)^2 . \] 
But $b |\EE_b| \leq 2 \len(G_n)$ and we get (\ref{pnupper}). 
For the other bound, write 
$\ell_i$ for the length of the longest edge at $x_i$.  
Then
\begin{equation}
 \Pr(\ell_I > a) \leq \frac{\Ex \ell_I}{a} 
= \frac{\sum_i \ell_i}{na} 
\leq \frac{2 \len(G_n)}{na} . \label{ell}
\end{equation}
Now observe that if a pair of vertices $(x_i,x_j)$ satisfies 
{\em  none} of the following three conditions \\
(i) $d(x_i,x_j) \leq 2a+b$ \\
(ii) $\max(\ell_i,\ell_j) > a$ \\ 
(iii) number of hops on route from $x_i$ to $x_j$ is $\geq 4$\\
then $(x_i,x_j)$ satisfies the criteria in the definition of $p_n(a,b)$. 
So 
\[ 1 - p_n(a,b) \leq \Pr(d(x_I,x_J) \leq 2a+b) 
+ \Pr(\max(\ell_I,\ell_J) > a) + \Pr(H_n \geq 4) . \] 
Applying the previous coupling construction to the first term on the right side, 
and applying (\ref{ell}) to the second term,
\[ 1 - p_n(a,b) \leq 2D_{s_n} + \Pr(d(U_1,U_2) \leq 2a+b+2s_n\sqrt{2}) 
+ \frac{4 \len(G_n)}{na}  + \Pr(H_n \geq 4) . \] 
By conditioning on $U_1$,
\begin{equation} \Pr(d(U_1,U_2) \leq r)  \leq \pi r^2/n \label{dUU} 
\end{equation}
giving (\ref{pnlower}).
\end{proof}

Returning to the proof of Theorem \ref{T1}(c), recall that we have a sequence of configurations 
$\bx^n$ such that $D_{s_n} \to 0$ for some $s_n = O(n^{3/10})$.
Apply Lemma \ref{L1} with 
$a_n = n^{3/10}$ and $b_n = \delta n^{1/2}$ and add the two inequalities; 
a brief calculation gives 
\[ 1 \leq (\sfrac{C}{\delta} +4) n^{-13/10} \len(G_n) + \pi \delta^2 + \Pr(H_n \geq 4) + o(1) \]
for a constant $C$. 
Now 
\begin{equation} \time(G_n) - \dist(\bx^n) \geq \Delta (\Ex H_n - 1) \label{tdD}
\end{equation}
and so by hypothesis (\ref{2+}) 
$\Ex H_n \leq 3 + o(1)$ and so 
$\Pr(H_n \geq 4)$ is bounded away from $1$ 
(and the same conclusion follows from the weaker assumption (\ref{3-})).  
So by choosing $\delta$ sufficiently small,
\[ n^{-13/10} \len(G_n) = \Omega(1) \]
as required.

\subsection{Proof of Proposition \ref{P2}}
\label{sec-p2}
(a) Consider as before the vertices $\TL(L,n)$ associated with the scaled triangular lattice.
Define a network $G_n$ to consist of all edges $(x_i,v)$ 
where $x_i \in \bx^n$ and $v \in \TL(L,n)$.
For each pair of vertices $(i,j)$ consider the shortest two-edge route;
it is easy to check that this route length is at most
$d(i,j) + 2c_3L$, and so
\[ \time(G_n) - \dist(\bx^n)  \leq \Delta + 2c_3L/n^{1/2} . 
\]
The number of edges of $G_n$ is 
$O(n \times n/L^2)$
and their lengths are $O(n^{1/2})$, so 
\[ \len(G_n) = O(n^{5/2}/L^2) . \]
Taking $L_n \sim \delta n^{1/2}$ for small $\delta$ establishes (a).

(b) 
Fix $r > 0$.
As before, write 
$\ell_i$ for the length of the longest edge at $x_i$.  
For each pair of vertices $(x_i,x_j)$ at least one of the following three conditions must hold.
\\
(i) $\max(\ell_i,\ell_j) \geq \sfrac{1}{2}rn^{1/2}$\\
(ii) $d(x_i,x_j) \leq rn^{1/2}$\\
(iii) the route from $x_i$ to $x_j$ has at least three hops.\\
So for random vertices 
$x_I,x_J$
\[ 2\Pr(\ell_I
\geq \sfrac{1}{2}rn^{1/2}) 
+ \Pr(d(x_I,x_J) \leq rn^{1/2})
+ \Pr(H_n \geq 3) \geq 1 \]
where $H_n$ is the number of hops on the route from $x_I$ to $x_J$.
Using (\ref{ell}) the first term is at most
$\frac{8 \len(G_n)}{rn^{3/2}}$.
Using (\ref{dUU}) 
and the equidistribution hypothesis, 
\begin{equation}
\Pr(d(x_I,x_J) \leq rn^{1/2})
\leq \pi r^2 + o(1).  \label{eqq}
\end{equation}
So
\[ 
\frac{8 \len(G_n)}{rn^{3/2}}
\geq 1 - \Pr(H_n \geq 3) - \pi r^2 - o(1) . \]
Using (\ref{tdD}) as before, hypothesis (\ref{2-}) 
implies that $\Pr(H_n \geq 3)$ is bounded away from $1$.
So for sufficiently small $r$
\[ \liminf_n 
\frac{8 \len(G_n)}{rn^{3/2}}
> 0 \]
establishing (b).

\subsection{Proof of Proposition \ref{P3}}
\label{sec-p3}
Part (a) follows from Proposition \ref{P2}(a) 
and general convergence arguments.
For (b) we use the following lemma,
proved by a simple compactness agument.
\begin{Lemma}
\label{L5}
For each $k \geq 1$ and $\rho > 0$ there exists $\eta(k,\rho) > 0$ such that the following holds.
Let $U$ be uniform on the unit square.
For each collection of points 
$(x; v_1,v_2,\ldots,v_k)$ 
in the unit square, and each function $0 \leq f(u) \leq 1$ with $\Ex f(U) \geq \rho$,
\[ \Ex \left[ \left(\min_{1\leq i \leq k} (d(x,v_i)+d(v_i,U) )
 - d(x,U) \right) \ f(U) \right]
\geq \eta(k,\rho)  . \]
\end{Lemma}
Set $\delta = \frac{1}{4} \pi^{-1/2}$ and let $k \geq 1$. 
Consider a network $G_n$ on vertex-set $\bx^n$.
Write $h(i,j)$ for the number of hops on the minimum-time route from $x_i$ to $x_j$. 
Define $\VV$ as the  set of $i$ such that at most $k$ edges at $x_i$ have lengths 
$\geq \delta n^{1/2}$, and for $i \in \VV$ let $\VV_i$ be the set of the other end-vertices of such edges. 
Define $B$ to be the set of pairs $(i,j)$ such that $h(i,j) = 2$ and the first hop on the route from 
$i$ to $j$ has length $\geq \delta n^{1/2}$.

Now consider a pair $i,j$ such that $i \in \VV$ and $(i,j) \in B$.
Let $(x_i,v,x_j)$ be the route from $i$ to $j$. 
So $v \in \VV_i$.
Using the triangle inequalities
\begin{eqnarray*}
d(v,u) &\leq& d(v,x_j) + d(x_j,u) \\
d(x_i,x_j) &\leq& d(x_i,u) + d(x_j,u)
\end{eqnarray*}
we find that, for any point $u$, 
\[ d(x_i,v) + d(v,u) - d(x_i,u) \leq 
d(x_i,v) + d(v,x_j) - d(x_i,x_j) 
+ 2d(x_j,u) . \]
Because 
$\time(i,j) = n^{-1/2}(d(x_i,v) + d(v,x_j) ) + \Delta$ 
we get 
\[ n^{-1/2}\min_{v \in \VV_i} (d(x_i,v) + d(v,u) - d(x_i,u)) \ 1_{(i \in \VV,(i,j) \in B)} 
\hspace*{0.7in} \] \[ \hspace*{0.7in} \leq \time(i,j) - \Delta - n^{-1/2}d(x_i,x_j) + 2n^{-1/2}d(x_j,u) . \] 
Apply this inequality to uniform random $I,J$ and the coupled pair 
$(x_J,U_n)$ given by Lemma \ref{L-couple}.
Taking expectations,
\begin{eqnarray*}
A_n&:=&
 n^{-1/2} \Ex \left[ \min_{v \in \VV_I} (d(x_I,v) + d(v,U_n) - d(x_I,U_n)) \ 1_{(I \in \VV,(I,J) \in B)}  \right] \\
&\leq& \time(G_n) - \Delta - \dist(\bx^n) + 2 n^{-1/2} \Ex d(x_j,U_n)
\end{eqnarray*} 
and the right side $\to 0$ as $n \to \infty$ by hypothesis (\ref{1-}) and by the 
coupling inequality.  But by conditioning on $I$ and applying Lemma \ref{L5}, 
\[ A_n \geq \Ex \eta(k,g(I)) 
\mbox{ where } 
g(i):= \Pr((i,J) \in B) \cdot 1_{(i \in \VV)} . \] 
As $n \to \infty$, the fact $A_n \to 0$ implies 
$\Ex g(I) \to 0$, that is
\begin{equation} 
\Pr((I,J) \in B, \ I \in \VV) \to 0 
\mbox{ as } n \to \infty  .
\label{PIJ}
\end{equation}
On the other hand, by symmetry 
$(I,J) \ed (J,I)$ 
and the fact that one hop of a two-hop route covering distance $ \geq 2\delta n^{1/2}$ must have 
length $ \geq \delta n^{1/2}$, 
\[ \Pr \left( (I,J) \in B \left| h(I,J) = 2, d(x_I,x_J) \geq 2\delta n^{1/2}, i \in \VV, j \in \VV \right. \right) \geq \sfrac{1}{2}. \] 
So 
\[ \Pr \left( (I,J) \in B , i \in  \VV \right) \geq \sfrac{1}{2}
\Pr \left( h(I,J) = 2, d(x_I,x_J) \geq 2\delta n^{1/2}, i \in \VV, j \in \VV  \right) . \] 
Now hypothesis (\ref{1-}) implies 
$\Ex h(I,J) \leq 2 + o(1)$.  
We must have $\Pr(h(I,J) = 1) \to 0$ because otherwise some non-vanishing proportion of all 
${n \choose 2}$ possible edges would appear in $G_n$, making $\len(G_n)$ be of order 
$n^{5/2}$. So
\[ \Pr(h(I,J) = 2) \to 1 . \] 
Using this and (\ref{eqq}), 
\[ \Pr \left( (I,J) \in B , i \in  \VV \right) \geq \sfrac{1}{2} - 4\pi \delta^2 - o(1) - 2\Pr(I \not\in \VV) . \] 
We chose $\delta$ to make $4\pi \delta^2 = \frac{1}{4}$, and now (\ref{PIJ}) implies
\[ \Pr(I \not\in \VV) \geq \sfrac{1}{8} - o(1) . \] 
The definition of $\VV$ gives the first inequality in 
\begin{eqnarray*}
\len(G_n) &\geq& \sfrac{1}{2}k \ \delta n^{1/2} \ |\VV^c|\\
&=& \sfrac{1}{2}k \ \delta n^{3/2} \Pr(I \not\in \VV) \\
&\geq& \delta k n^{3/2} (\sfrac{1}{16} - o(1)).
\end{eqnarray*} 
Because $k$ is arbitrary, we have proved 
$n^{-3/2} \len(G_n) \to \infty$.

\subsection{Proof of Theorem \ref{T4}} 
\label{sec-PT4}
(a) Let $x_*$ be the center of the square $[0,n^{1/2}]^2$. 
It is enough to construct networks such that the maximum hop-distance from any vertex to $x_*$ 
is $\leq (1 + o(1)) \log_2 \log n$.
We do this via a hierarchical construction of hubs.  
Define 
\[ h_n  =   \log_2 \log n  \] 
and then take $b_n$ such that $b_n/h_n \to \infty$ and $\frac{\log b_n}{h_n} \to 0$.  
Define
\begin{eqnarray*}
s(n,0)&=& n^{1/2}\\
m(n,j) &=&\lfloor  \sqrt{s(n,j-1)/h_n} \rfloor \\
s(n,j) &=& s(n,j-1)/m(n,j), \ 1 \leq j \leq J_n
\end{eqnarray*} 
for 
\[ J_n:= \min \{j: s(n,j) \leq b_n\} . \] 
Then define 
\[ s(n,j) = \sfrac{1}{2} s(n,j-1), \ J_n +1 \leq j \leq K_n \]
for
\[ K_n:= \min \{j: s(n,j) \leq 1\} . \] 
Let $\SS_j$ be the natural partition of $[0,n^{1/2}]^2$ into subsquares $\sigma$ 
of side $s(n,j)$, and write $x_\sigma$ for the center of the subsquare $\sigma$. 
Given an arbitrary configuration $\bx^n$, define the edges of $G_n$ to be \\
(i) for each $1 \leq j \leq K_n$ and each $\sigma \in \SS_j$, the edge 
$(x_\sigma,x_{\hat{\sigma}})$, 
where $\hat{\sigma} \in \SS_{j-1}$ denotes the supersquare containing $\sigma$. \\
(ii) for each $x \in \bx^n$, the edge $(x,x_\sigma)$, where $\sigma \in \SS_{K_n}$ is the subsquare containing $x$ in the finest partition. \\
So the natural route from $x$ to $x_*$ uses $1 + K_n$ hops.   
Ignoring the ``integer part" in the definition of
$m(n,j)$ 
(it is easy to check this makes no difference to the asymptotics below),
\[ s(n,j) = s(n,j-1)/m(n,j) = \sqrt{s(n,j-1) h_n} \]
which is
\[ \log s(n,j) = \sfrac{1}{2} \log s(n,j-1) + \sfrac{1}{2}\log h_n \]
and so 
\[ \log s(n,J_n) \leq 2^{-J_n} \log s(n,0) + \log h_n . \] 
If $J_n \geq h_n$ then the right side would be $O(\log h_n)$, contradicting the definition of $J_n$, 
and so we must have $J_n \leq h_n$ ultimately.  
Noting that $K_n - J_n \leq 1 + \log_2 b_n$ we have shown $K_n \leq (1 + o(1)) \log_2 \log n$, 
establishing the desired bound on hop-distance.

To bound the network length, we have 
\[ \len(G_n) \leq  \sqrt{2} + \sum_{j=1}^{K_n} \frac{n}{s^2(n,j)} \times \sqrt{2} s(n,j-1) \] 
and so 
\begin{eqnarray*}
\frac{\len(G_n)}{n \sqrt{2}} - 1 &\leq& \sum_{j=1}^{K_n} \frac{s(n,j-1)}{s^2(n,j)} \\
&=&  \sum_{j=1}^{J_n} \frac{m^2(n,j)}{s(n,j-1)} + \sum_{j=J_n +1}^{K_n} \frac{2}{s(n,j)} \\
&\leq& J_n/h_n + 8
\end{eqnarray*} 
using the definition of $m(n,j)$ for the first term and the fact $s(n,K_n) \geq 1/2$ for the second term. 
Since $J_n \leq h_n$ ultimately, we have proved $\len(G_n) = O(n)$.

(b) Fix $n$ and  a network $G_n$ on vertex-set $\bx^n$.  
From the definition of $\hop(G_n)$ as an average, there exists some distinguished 
vertex $v^*$ 
such that 
\[ \ave_i \  \hop(i,v^*) \leq \hop(G_n). \] 

{\em  Remark.}  The analysis below largely follows the idea in (a); a minimal-hop network for distributing material from a ``center" $v^*$ to other vertices must be somewhat like the hierarchical scheme in (a).  
In particular it uses longest edges in the first step.  Of course, in the setting of $\hop(G_n)$ this cannot be true for {\em every} initial vertex $v$, and this is where the argument is inefficient.

Consider an array of integers 
\[ m(n,j) \geq 4; \quad 0 \leq j \leq J_n \] 
which will be defined precisely later.  
Use this array to define
\[ s(n,0) = n^{1/2}/m(n,0) \]
and inductively for $1 \leq j \leq J_n$
\[ s(n,j) = s(n,j-1)/m(n,j) . \] 
Let $\SS_j$ be the natural partition of $[0,n^{1/2}]^2$ into subsquares $\sigma$ 
of side $s(n,j)$.  
We will bound (Lemma \ref{L13} below) hop-distance from $v^*$ in terms of the following construction which we envisage as an {\em infection process}.  
In this process, we define subsets $\II_j \subset \SS_j$ of squares $\sigma$ be be infected 
(either ``primary" or ``secondary" or both) according to the following rules.

\noindent
(i) $\II_0$ consists of the square $\sigma^* \in \SS_0$ containing the distinguished 
vertex $v^*$.

\noindent
Now for $\sigma \in \SS_j$ write $\hat{\sigma} \in \SS_{j-1}$ for the supersquare containing $\sigma$.  
For $1 \leq j \leq J_n$:

\noindent
(ii) $\sigma \in \SS_j$ is infected as a secondary infection if $\hat{\sigma}$ or a square of $\SS_{j-1}$ adjacent to $\hat{\sigma}$ is infected;\\
(iii) $\sigma \in \SS_j$ is infected as a primary infection if there exists an edge of length 
$> s(n,j-1)$ with one end-vertex in $\sigma$.

\noindent
(In (ii), ``adjacent" includes diagonally adjacent).
Write $\II_j$ for the set of infected squares of $\SS_j$.
\begin{Lemma}
\label{L13}
Define $\RR_j \subset \SS_j$ to be the set of squares $\sigma$ such that there exists a route 
from $v^*$ to some vertex $x \in \sigma$ with at most $j$ hops. 
Then $\RR_j \subset \II_j$.
\end{Lemma}
\begin{proof}
$\RR_0 = \II_0 = \{\sigma^*\}$, so suppose inductively that $\RR_{j-1} \subset \II_{j-1}$.  
Let $\sigma \in \RR_j$, and consider the first edge $e = (x,v)$ on a route from some $x$ to $v^*$ with at most $j$ hops.
In the case $\len(e) > s(n,j-1)$, then $\sigma$ is infected by (iii).  
Write $\sigma^\prime \in \SS_{j-1}$ for the square containing $v^*$. 
So $\sigma^\prime \in \RR_{j-1} \subset \II_{j-1}$ is infected, and in the alternate case 
$\len(e) \leq s(n,j-1)$ we have that $\hat{\sigma}$ is the same as or adjacent to $\sigma^\prime$, 
so $\sigma$ is infected by (ii). 
\end{proof}

\setlength{\unitlength}{0.05in}
\begin{picture}(37,39)(-21,0)
\multiput(4,4)(0,9){4}{\line(1,0){27}}
\multiput(4,4)(9,0){4}{\line(0,1){27}}
\put(1,1){\line(1,0){33}}
\put(1,34){\line(1,0){33}}
\put(1,1){\line(0,1){33}}
\put(34,1){\line(0,1){33}}
\multiput(1,4)(0,3){10}{\line(1,0){3}}
\multiput(31,4)(0,3){10}{\line(1,0){3}}
\multiput(4,1)(3,0){10}{\line(0,1){3}}
\multiput(4,31)(3,0){10}{\line(0,1){3}}
\multiput(0,0)(0,1){36}{\line(1,0){1}}
\multiput(34,0)(0,1){36}{\line(1,0){1}}
\multiput(0,0)(1,0){36}{\line(0,1){1}}
\multiput(0,34)(1,0){36}{\line(0,1){1}}
\put(0,0){\line(1,0){35}}
\put(0,35){\line(1,0){35}}
\put(0,0){\line(0,1){35}}
\put(35,0){\line(0,1){35}}
\put(16,16){$\sigma^\dagger$}
\end{picture}

\vspace{0.1in}
{\bf Figure 2.}
{\small 
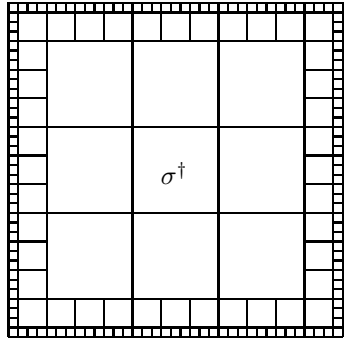
Upper bounding the spread of infection from a
primary infection of square $\sigma^\dagger$.
For visual clarity we draw the case $m(n,j) = 3$.
}

\vspace{0.1in}
\noindent
Now for any infected square of $\SS_k$, the infection can be traced back to some primary infection of some square of $\SS_j$ for some $0 \leq j \leq k$.
Conversely, each primary infection of some $\sigma^\dagger \in \SS_j$ creates an epidemic of infection in smaller squares.  It is geometrically clear (see Figure 2) that the total area of the squares thus infected in $\SS_{J_n}$ is bounded by $C_1\  \area (\sigma^\dagger)$ for some constant $C_1$.

Writing $A_n$ for the total area of infected squares of $\SS_{J_n}$, we see 
\[ A_n \leq C_1 \sum_{j=0}^{J_n} s^2(n,j) \times  |\{\sigma \in \SS_j: \sigma \mbox{ has primary infection } \}|  . \]
Because a primary infection in $\SS_j$ is caused by an edge of length $> s(n,j-1)$,
for $j \geq 1$
\[ |\{\sigma \in \SS_j: \sigma \mbox{ has primary infection } \}| \leq \frac{2 \ \len(G_n)}{s(n,j-1)} . \]
So
\begin{eqnarray*}
 A_n & \leq & C_1 \frac{n}{m^2(n,0)} +  2 C_1 \len(G_n) \ \sum_{j=1}^{J_n} \frac{s^2(n,j)}{s(n,j-1)} 
\\ 
&=& C_1 \frac{n}{m^2(n,0)} +     2 C_1 \len(G_n) \ \sum_{j=1}^{J_n} \frac{s(n,j-1)} {m^2(n,j)}.   
\end{eqnarray*}
Now suppose,
for some $\delta < 1/2$,
\begin{equation}
 A_n \leq \delta \mbox{ and } D_{s(n,J_n)}(\bx^n) \leq \delta . \label{AnD}
 \end{equation} 
Then at least $n(1 - 2 \delta)$ vertices of $\bx^n$ are not in infected squares of $\SS_{J_n}$, and so by 
Lemma \ref{L13} these vertices are at hop-distance at least $J_n + 1$ from $v^*$, implying
\begin{equation}
 \hop(G_n) \geq (1 - 2 \delta) (J_n + 1) . \label{GJ}
\end{equation}
Now fix $0 < \alpha < 1 - \beta$ 
and a constant $B$.
Define
\begin{eqnarray*}
J_n &=& \alpha \log_2 \log n
\\
h_n &=& B J_n
\\
m(n,0) &=& \lceil B C_1 \rceil
\\
m(n,j) &=& \lceil \sqrt{h_n \ s(n,j-1)} \rceil .
\end{eqnarray*}
So
\[ \frac{s(n,j-1)}{m^2(n,j)} \leq \frac{1}{h_n}, \quad j \geq 1 \]
implying 
\[ \frac{A_n}{n} \leq \frac{1}{B^2} 
+ 2C_1  \ \frac{\len(G_n)}{n} \ \frac{J_n}{h_n} 
= \frac{1}{B^2}
+ \frac{2C_1}{B}  \ \frac{\len(G_n)}{n} . \] %\frac{J_n}{h_n} 
Ignoring the ``next integer" in the definition of
$m(n,j)$ 
(it is easy to check this makes no difference to the asymptotics below),
\[ s(n,j) = s(n,j-1)/m(n,j) = \sqrt{s(n,j-1)/h_n} \]
which is
\[ \log s(n,j) = \sfrac{1}{2} \log s(n,j-1) - \sfrac{1}{2}\log h_n \]
and so 
\[ \log s(n,J_n) \geq 2^{-J_n} \log s(n,0) - \log h_n . \]
Using the definition of $J_n$
\[ \log_2 \log s(n,J_n) \geq (1 - \alpha - o(1)) \log_2 \log n . \]
Now the sequence $(s_n)$ in the hypothesis of (b) has 
\[ \log_2 \log s_n \leq 
(\beta + o(1)) \log_2 \log n \]
and so by choice of $\alpha$ we have 
$s(n,J_n)/s_n \to \infty$.
By hypothesis 
$D_{s_n}(\bx^n) \to 0$ 
and so under ``greater smoothing" we have also
\begin{equation}
D_{s(n,J_n)}(\bx^n) \to 0 .
\label{DsJ}
\end{equation}
Because $\len(G_n) = O(n)$, for given $\delta > 0$
we can choose $B$ such that $n^{-1}A_n \leq \delta$ ultimately.
Now the assumptions (\ref{AnD}) are satisfied, and (\ref{GJ}) implies
$ \hop(G_n) \geq (1 - 2 \delta) (J_n + 1) $.  %  \label{GJ}
Since $\delta$ is arbitrary and we can choose $\alpha$ arbitrarily close to 
$1 - \beta$
we have established (b).

%\newpage
\section{Final remarks}
\subsection{Geometry of near-optimal networks}
In the setting of Theorem \ref{T1}
($3$-hop networks)
we speculate that under equidistribution assumptions
there is a certain ``rigidity" property, 
that every near-optimal network has the same general features 
-- a highly-connected ``core"  
of about $n^{2/5}$ hubs.
On the other hand, in the setting of $2$-hop networks 
(Proposition \ref{P3})
our construction is clearly not optimal in detail 
(e.g. because it provides two possible routes $i \to j$, via a hub near $i$ or a hub near $j$, so one could construct more efficient networks containing only one of these alternatives).
So in this setting we speculate that different near-optimal networks can have
different statistical properties.

\subsection{Average $m$-hop networks}
\label{sec-mhop}
Theorem \ref{T1} and Proposition \ref{P3} suggest that, if we constrain the average number of hops to be some integer $m \geq 2$, the minimum network length will scale as 
$n^{\theta(m)}$ (possibly with an ``$\omega_n \to \infty$ arbitrarily slowly" term) for some exponent $\theta(m)$, 
where we know
\[ \theta(2) = \sfrac{3}{2}, \quad \theta(3) = \sfrac{13}{10} . \]  
The following recursive behavior for $m \geq 4$ seems intuitively clear, though we will not attempt to give details.
The optimal network has hubs at a certain density $n^{- \gamma(m)}$.  
Each point of $\bx^n$ is connected to its nearest hub.
The hubs are connected as in the optimal networks for average $m-2$ hops.  
By calculating the length of such a network and optimizing over $\gamma(m)$ we get a recursion
\[ \theta(m) = \frac{3 \theta(m-2) - 1}{2 \theta(m-2)} \] 
whose solution is
\[ \theta(2m) = 1 + \sfrac{1}{2^{m+1} - 2}, \quad 
\theta(2m+1) = 1 + \sfrac{3}{2^{m+3} - 6} \]
so that 
\[ \theta(4) = \sfrac{7}{6}, \ 
\theta(5) = \sfrac{29}{26}, \  
\theta(6) = \sfrac{15}{14}, \ 
\theta(7) = \sfrac{61}{58}  \ldots \ldots . \] 
Note also that the formula gives  $\theta(1) = \frac{5}{2}$, corresponding to the length $O(n^{5/2})$ of the compete graph on all $n$ vertices.

\subsection{Hierarchical networks over random points}
Hierarchical networks over random points have been studied in other contexts. 
For instance \cite{BKLZ,ZDR} emphasize the use of standard point process methods for finding exact formulas within such constructions built over Voronoi cells.

 %\newpage
 %\bibliographystyle{plain}
 %\bibliography{../../trees/me,../../trees/networks,../../trees/alg,../../trees/misc}

 \end{document}